\documentclass[aps,
               prl,
               twocolumn,
               nofootinbib,
               floatfix,
               showpacs
              ]{revtex4}

\usepackage{amsmath, amssymb}
\usepackage{graphics}
\usepackage{pstricks}
\usepackage{graphicx}
\usepackage{color}
\definecolor{darkblue}{rgb}{0.0,0.0,0.4}
\definecolor{darkgreen}{rgb}{0.0,0.4,0.0}
\definecolor{darkred}{rgb}{0.6,0.0,0.0}
\usepackage[bookmarksopen]{hyperref}
\hypersetup{colorlinks,linkcolor=darkgreen,citecolor=darkblue,urlcolor=darkblue}

\renewcommand{\i}{\mathrm{i}}

\renewcommand{\vector}[1]{\mathbf{#1}}

\newcommand{\del}{\partial}

\newcommand{\Fig}[1]{Fig.~\ref{fig:#1}}

\begin{document}

\title{Universal dynamics on the way to thermalisation}

\author{Boris Nowak}
\author{Jan Schole}
\author{Thomas~Gasenzer}
\affiliation{Institut f\"ur Theoretische Physik,
             Ruprecht-Karls-Universit\"at Heidelberg,
             Philosophenweg~16,
             69120~Heidelberg, Germany}
\affiliation{ExtreMe Matter Institute EMMI,
             GSI Helmholtzzentrum f\"ur Schwerionenforschung GmbH, 
             Planckstra\ss e~1, 
             64291~Darmstadt, Germany} 

\date{\today}

\begin{abstract}
It is demonstrated how a many-body system far from thermal equilibrium can exhibit universal dynamics in passing a non-thermal fixed point.
As an example, the process of Bose-Einstein (BE) condensation of a dilute cold gas is considered.
If the particle flux into the low-energy modes, induced, e.g., by a cooling quench, is sufficiently strong, the Bose gas develops a characteristic power-law single-particle spectrum $n(k)\sim k^{-5}$, and critical slowing down in time occurs.
The fixed point is shown to be marked by the creation and dilution of tangled vortex lines.
Alternatively, for a weak cooling quench and particle flux, the condensation process runs quasi adiabatically, passing by the fixed point in far distance, and signatures of critical scaling remain absent.
\end{abstract}


\maketitle

\textit{Introduction.} 
Systems like a ferromagnet or a bosonic quantum gas can undergo a phase transition of second order in the Ehrenfest classification, at which their relevant physical properties become independent of microscopic details. 
This independence leads to the concept of universality which has become extremely successful in classifying and characterizing matter in thermal equilibrium.
Many different phenomena can be characterized in terms of a few classes governed by the same critical properties.
Generalizing this concept to universal dynamics far away from thermal equilibrium requires to develop an understanding of the structure of the space of  possible non-thermal states.
In addition to the well-known dynamics near thermal fixed points \cite{Hohenberg1977a} critical regions, such as fixed points, critical lines and surfaces can affect the evolution far from equilibrium. 
When a closed system approaches such critical configurations, memory about the particular initial state it has started from would be partially lost.
Critical slowing down in the actual time evolution would be observed since the largest scales, not only in space but also in time, dominate the systems dynamics.
Thermal states are attractive, stable fixed points in this framework. 
Different systems would be related by means of the universality class their dynamical evolution falls into. 
Predictions for the behavior of very different physical systems could be obtained on the basis of comparatively few exemplary measurements. 
When looking at fundamental science applications, this could link the dynamics observed in very different contexts.

Here we show that such universal time evolution is possible, at the example of a strongly shock-cooled three-dimensional normal-fluid Bose gas which approaches a non-thermal fixed point \cite{Berges:2008wm,Bonini1999a} before it eventually proceeds to a thermal distribution below the Bose-Einstein critical temperature.
The cooling quench puts the system far out of equilibrium such that it can undergo an evolution passing the vicinity of a critical point where it gets stuck for a long time due to a critically enhanced role of ultra low-energy modes.
Looking at the critical configuration in real space we find it to be dominated by a dilute ensemble of quasi-topological vortex line defects which only very slowly decay via interactions with sound.

\begin{figure}[!t]
\ \\[-6ex]\includegraphics[width=0.35\textwidth]{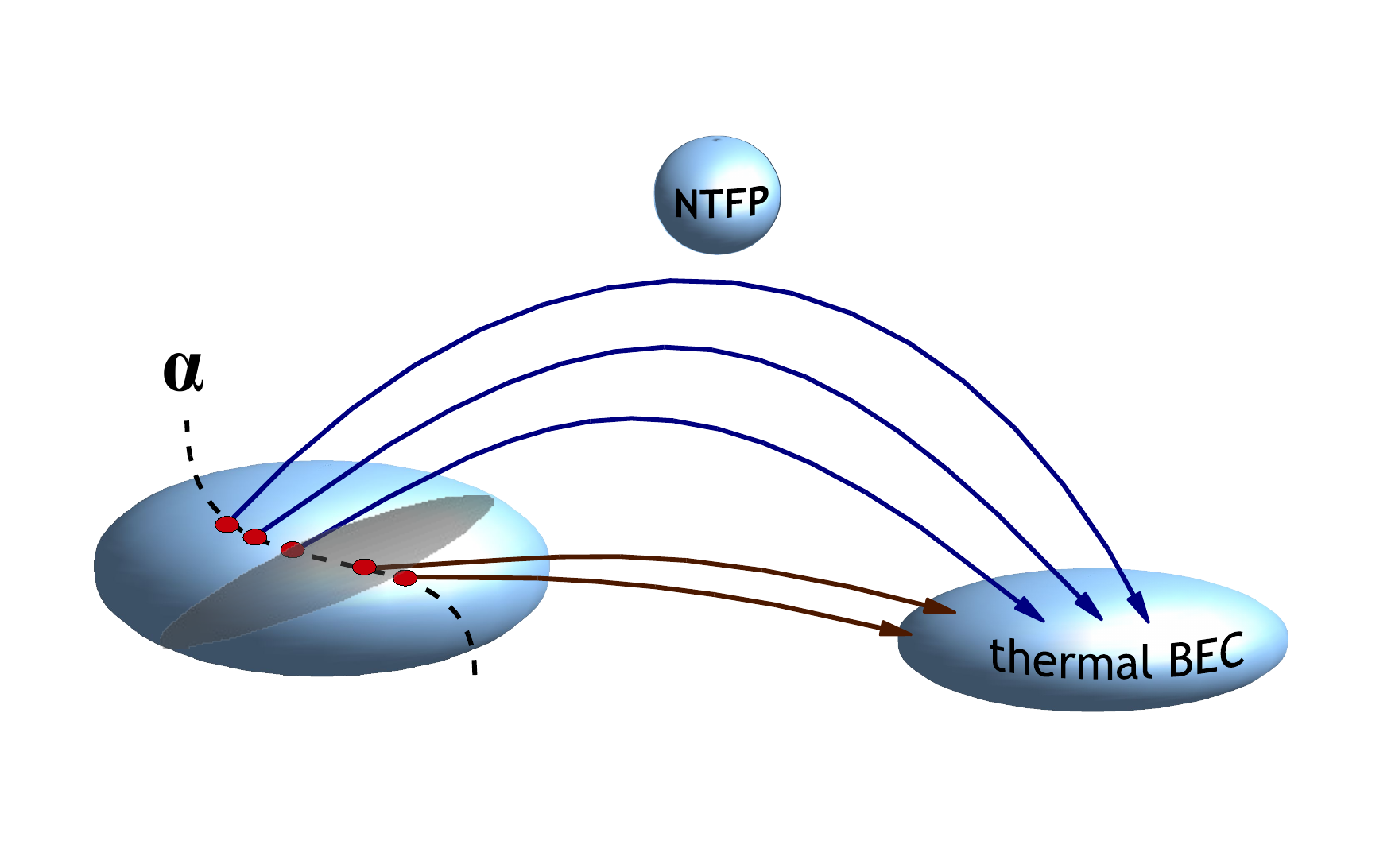}
\ \\[-6ex]
\caption{(Color online) Depending on the strength $\alpha$ of an initial cooling quench, the gas can thermalize directly to a Bose-Einstein condensate, or it can first approach and critically slow down near a non-thermal fixed point (NTFP). 
Thereby it is characterized by a self-similar particle spectrum $n(k)\sim k^{-5}$ and a diluting ensemble of growing-size vortex rings.
}
\label{fig:NTFPScheme}
\end{figure}
The formation of a Bose-Einstein condensate (BEC) from a disordered state has been the subject of many studies~\cite{Levich1977a, Stoof1991a, Svistunov1991a, Kagan1992a, Kagan1994a, 
Damle1996a, Semikoz1997a, 
Kagan1997c, Snoke1989a, Gardiner1998b, Miesner1998a, Drummond1999a, 
Kohl2002a, Berloff2002a, Connaughton2005a, Nazarenko2006a, Ritter2007a, Weiler2008a, Kozik2009a,Smith2011a}. 
Among other aspects, the role of superfluid turbulence in this process has been discussed~\cite{Svistunov1991a,Kagan1992a, Kagan1994a, 
Berloff2002a, Nazarenko2006a}. 
Bose-Einstein condensation in a non-equilibrium under-cooled gas can have the characteristics of a turbulent inverse cascade \cite{Svistunov1991a,Kagan1992a, 
Berloff2002a}.

In this article, we demonstrate how the superfluid turbulence period is set into the context of universal dynamics.
Depending on whether the condensation evolution passes close to or further away from a non-thermal fixed point, the process appears in qualitatively different forms  (\Fig{NTFPScheme}).
Suppose we start with a cold dilute, incoherent homogeneous gas in three dimensions, at a temperature above the BEC phase transition. 
Let us apply a cooling quench to this gas by removing particles from the high-momentum tail of the distribution and suppose that,
appropriate collisions provided, the system re-equilibrates to thermal equilibrium.
The final temperature is then given by the total energy after the quench which we assume to be below the BEC critical temperature. 
We remark that the question whether a system can thermalize at all \cite{Gogolin2011a} can be addressed as the question of which types of stable attractors exist.

\begin{figure}[!t]
\includegraphics[width=0.35\textwidth]{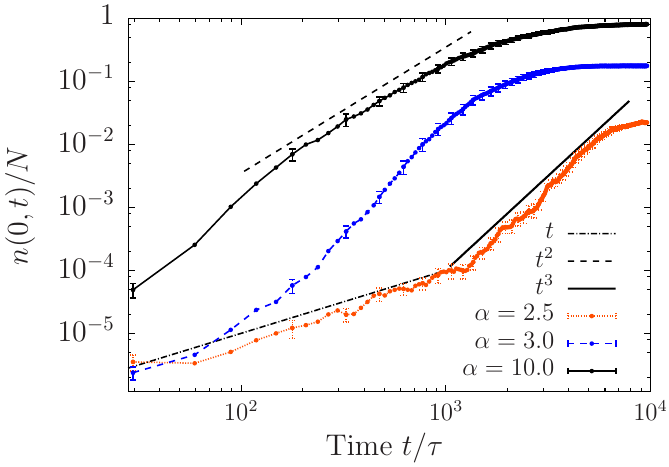}
\\[-2ex]
\caption{(Color online) Build-up of the condensate fraction $n(\mathbf{k}=0,t)/N$ on a double-logarithmic scale, with $\tau=m\xi^{2}$, for different strengths of the initial cooling quench, parametrized by $\alpha$.
Depending on $\alpha$, different power-law behavior is seen. Standard averaging errors are shown.
}
\label{fig:ZeroMode}
\end{figure}
Two different scenarios are possible: If a sufficiently small amount of energy is removed, the subsequent scattering of particles into the low-energy modes builds up a thermal Rayleigh-Jeans distribution $n(k)\sim k_\mathrm{B}T/k^{2}$ in a quasi-adiabatic way. 
The chemical potential increases, and a fraction of particles is deposited in the lowest mode, forming a BEC. 
During this process, tangles of defect lines can be found in the Bose field by filtering out short-wavelength fluctuations \cite{Berloff2002a}. 

In the second scenario, given a sufficiently strong cooling quench cutting away, e.g., all particles above a cutoff $k_{c}$, the remaining overpopulation of the modes just below $k_{c}$ results in a vigorous transport towards lower energies that has the form of a strong-wave-turbulence inverse cascade  \cite{Nowak:2011sk}.
This cascade induces a long-lived, power-law single-particle spectrum $n(k)\sim k^{-\zeta}$, with an exponent $\zeta=5$ distinctly larger than the exponent $\zeta=2$ of the thermal Rayleigh-Jeans distribution. 
The emergence of the strong cascade can be explained by the dominance of vortical superfluid flow around line defects over compressible, longitudinal sound excitations and density fluctuations in the respective regime of wave lengths.
The power law $k^{-5}$ signals that the system evolves close by a distinct non-thermal fixed point (NTFP) \cite{Berges:2008wm,Berges:2008sr,Scheppach:2009wu} where the evolution critically slows down. 
The particular power can be traced back to the flow pattern around the vortex lines \cite{Nowak:2011sk} which now become visible without filtering out short-wavelength fluctuations. 
The two possible paths to BE condensation are shown schematically in \Fig{NTFPScheme}.
Whether the system, during the condensation dynamics, can approach the NTFP or moves in a direct way to thermal equilibrium depends, for a closed system, on the initial conditions, i.e., on the strength of the cooling quench.
We refer to the condensation process which takes the `detour' via the NTFP as \emph{hydrodynamic BE condensation} because of the dynamical scale separation of incompressible and compressible components of the particle current.

\begin{figure}[!t]
\includegraphics[width=0.37\textwidth]{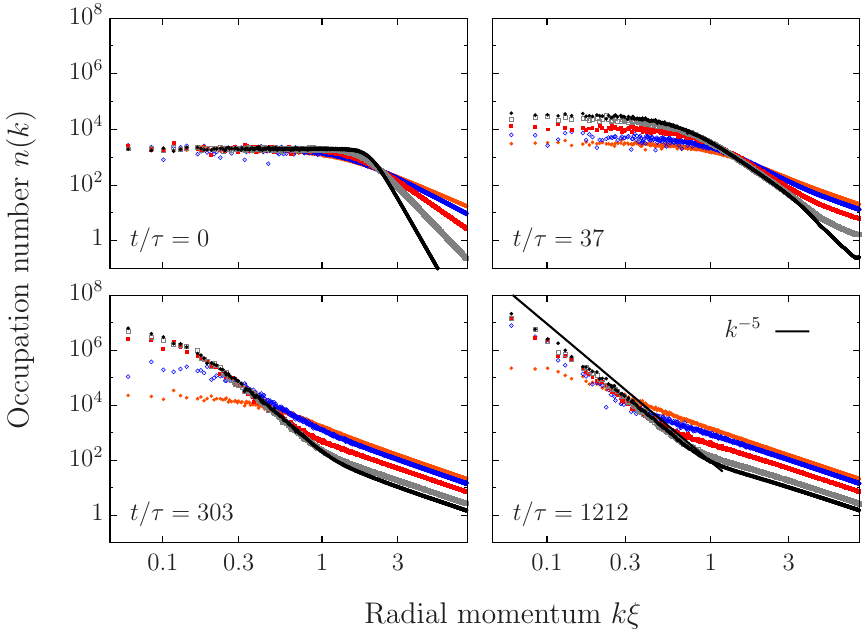}
\caption{(Color online) Particle momentum spectra at four different times, on a double-logarithmic scale as functions of the radial momentum $k=|\mathbf{k}|$ for different $\alpha=2.5$ (orange), $3.0$ (blue), $4.0$ (red), $6.0$ (grey), $10$ (black), from top to bottom at $k\xi=1$. $\tau=m\xi^{2}$.
Averages over $3..7$ runs. 
A steep infrared power-law $n(k)\sim k^{-5}$ appears for $\alpha>3$.
}
\label{fig:SpectrumEvolution}
\end{figure}

\textit{Semiclassical simulations.}
To reveal this dynamics we study a dilute Bose gas, in the classical-wave limit, sampling statistically the classical equation of motion which has the form of the Gross-Pitaevskii equation (GPE) for the classical field $\phi(\mathbf{x},t)$, see the supplementary material \cite{Supplement} for details.
We compute the evolution in a cubic volume with periodic boundary conditions.
The initial field in momentum space, $ \phi(\mathbf{k},0)$ is parametrized such that its phase is random while the modulus squared gives a momentum distribution $n(\mathbf{k},0)$ which is flat between $k=|\mathbf{k}|=0$ and $k\simeq k_{0}$ while it falls off as $k^{-\alpha}$ for $k\gtrsim k_{0}$ as shown in the top left panel of \Fig{SpectrumEvolution}.

\textit{Evolution of the zero mode.}
We consider an initial power-law fall-off of $n(k)$  close to or steeper than that expected by a self-similar solution of the wave Boltzmann equation, $\alpha\simeq2.4$ \cite{Semikoz1997a}, corresponding to an inverse particle cascade in weak wave turbulence theory \cite{Svistunov1991a,Kagan1992a}.
$\alpha=2$ would correspond to a stable initial thermal distribution at finite chemical potential, and thus any $\alpha>2$ is required to set off a re-equilibration to a lower temperature.
In \Fig{ZeroMode} we show the ensuing time evolution of the condensate occupation number $n(0,t)$ for the different $\alpha$. 
In each case, the evolution leads to a BEC characterized by a non-vanishing ratio $n(0,t)/N$. 
As we keep $n(k=0,0)$ and $k_{0}^{\alpha}$ constant $n(0,t)/N$ grows with $\alpha$. 
This is because larger $\alpha$ cut off more high-momentum particles and leave less energy to be thermally redistributed.
Power-law growth $n(0,t)\sim t$ predicted in \cite{Svistunov1991a}, and later $\sim t^{3}$ \cite{Damle1996a} is seen for near-thermal $\alpha=2.5$, while for large $\alpha$ the late-time growth reduces to $\sim t^{2}$.
As a result, strong cooling quenches qualitatively modify the way how the BEC grows, and a slowing down of this process is seen.

\textit{Momentum distribution.}
Beyond the zero mode, the occupation spectrum of the non-zero momentum modes allows to follow the dynamical process of strongly wave-turbulent particle transport to lower energies and to identify a further smoking gun for the approach of the non-thermal fixed point.
In \Fig{SpectrumEvolution} we show the time evolution of the single-particle distribution $n(k,t)$ over the absolute momenta $k=|\mathbf{k}|$ at four different times and for different $\alpha$. 
During the initial evolution ($t \lesssim 10^2\tau$) many particles gradually move to lower wave numbers, while at the same time relatively few particles deposit the surplus energy in the high-momentum modes, refilling them again. 
According to Ref.~\cite{Semikoz1997a}, a weak-wave-turbulence inverse particle cascade with $n(k)\sim k^{-2.4}$ is expected within the range of validity of  kinetic theory.
After a while, however, the spectra developing from the different initial $\alpha$ differ strongly. 
For $\alpha\gtrsim3$, the distribution develops a bimodal structure, with $n(k) \sim k^{-5}$ in the infrared (IR) and $n(k) \sim k^{-2}$ in the UV. 
At very long times, this bimodal structure decays towards a global $n(k) \sim k^{-2}$ (not shown). 
For $\alpha \lesssim 3$, the distribution directly reaches a thermal Rayleigh-Jeans scaling $n(k)\sim T/k^{2}$. 
%

\textit{Hydrodynamic condensation.}
To interpret our results in the context of non-thermal fixed points we analyse kinetic-energy spectra as in \cite{Nore1997a}, see \cite{Supplement} for details. 
In \Fig{IncompSpectrum1}a, we show the evolution of the different components $n_{\delta}(k)$, $\delta\in\{$i, c, q$\}$, for the weak initial quench $\alpha=2.5$: 
The incompressible component $n_{i}$ which accounts for vortical flow, with a velocity vector changing transversally to its direction, is of roughly equal magnitude as the compressible component $n_{c}$ of the longitudinal density fluctuations, i.e., sound excitations.
At the same time, the quantum pressure component $n_{q}$ is insignificant on all scales. 
At early times, $t\lesssim10^{3}\tau$, due to the absence of phase coherence \cite{Nowak:2011sk} the resulting spectra  do not add up to the single-particle spectrum $n(k)\not=n_\mathrm{i}(k)+n_\mathrm{c}(k)+n_\mathrm{q}(k)$. 
$n(k)$ grows in the regime of low momenta meaning that phase coherence is being established and growing, and a condensate fraction appears (\Fig{ZeroMode}). 
For the case of the strongly non-thermal initial distribution, $\alpha=10$, the evolution is shown in \Fig{IncompSpectrum5}b. 
In contrast to \Fig{IncompSpectrum1}a, two macroscopic flows can be observed, one to the UV, and one to the IR.
Conservation of particle and energy imply immediately that, when sent out from the regime of intermediate frequencies, energy is deposited in the UV while particles are predominantly transferred to the IR. 
This leads to an inverse particle cascade with approximately $k$-independent radial particle flux $Q(k)\equiv Q$ and a corresponding direct energy cascade to the UV \cite{Nowak:2011sk}.
The inverse particle cascade reflects strong wave turbulence, characterized by $n(k)\sim k^{-5}$ \cite{Scheppach:2009wu}. 
The decomposition in \Fig{IncompSpectrum1}b makes clear that this power-law is caused by incompressible excitations only \cite{Nowak:2010tm,Nowak:2011sk}, establishing a dominantly ideal \emph{hydrodynamic} (superfluid) BE condensation process. 
In the UV, the excitations follow a thermal $n(k)\sim k^{-2}$ and are dominated by $n_{c}$ and $n_{q}$. 

\begin{figure}[!t]
\flushleft
(a)\hfill\ \\[-3ex]
\hspace*{0.017\textwidth}
\includegraphics[width=0.4\textwidth]{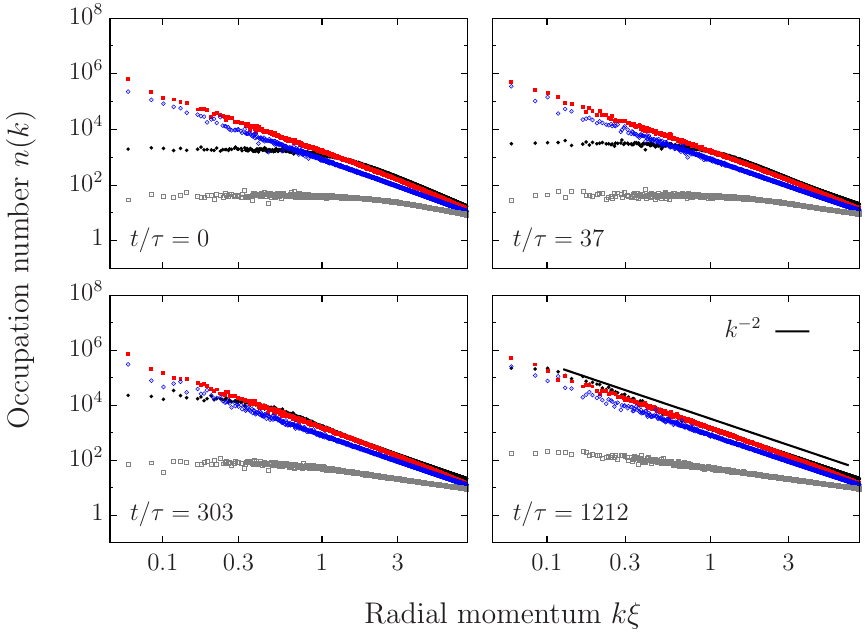}
\ \\
(b)\hfill\ \\[-3ex]
\hspace*{0.017\textwidth}
\includegraphics[width=0.4\textwidth]{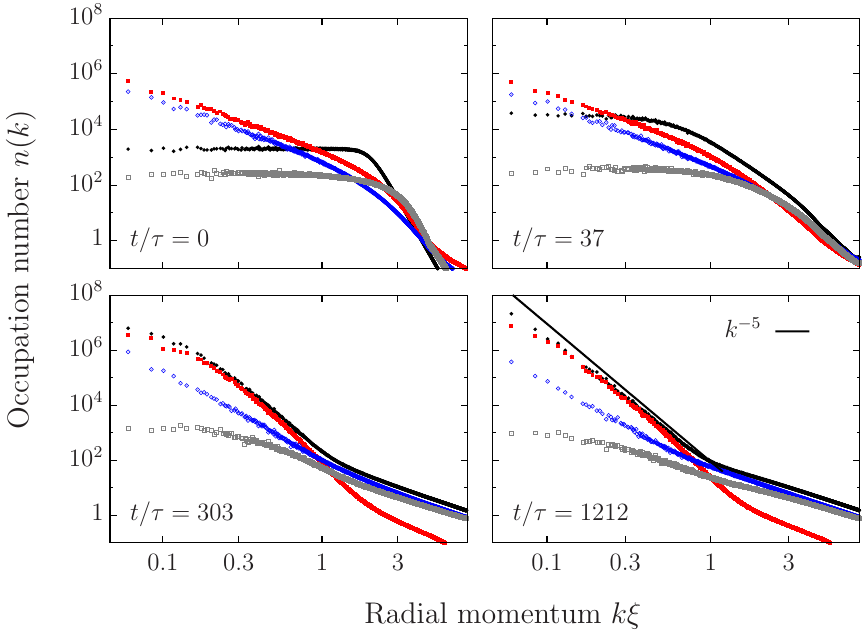}
\caption{(Color online) Decomposition of particle spectra $n(k)$ (black) into incompressible (red), compressible (blue), and `quantum-pressure' (grey) components, see text for definitions, on a double-log scale, at four different times, in units of $\tau=m\xi^{2}$.
(a) Weak cooling quench, $\alpha=2.5$. Average of 3 runs (b) Strong quench, $\alpha=10$. Average of 7 runs.
}
\label{fig:IncompSpectrum1}
\label{fig:IncompSpectrum5}
\end{figure}
%
The above results show that during the hydrodynamic condensation process incompressible flow temporarily dominates in the IR regime at the expense of compressible excitations.
The opposite occurs for the compressible excitations in the UV.
This dynamical scale separation and bimodal power-law distribution is a signature of the system approaching the NTFP.

\begin{figure}[!t]
 \includegraphics[width=0.37\textwidth]{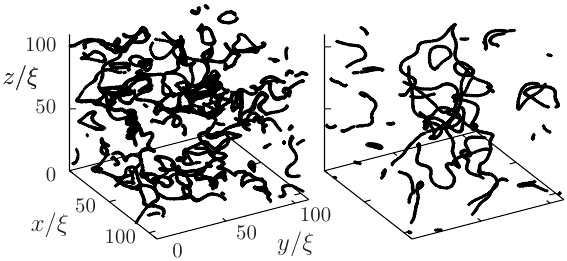}
 \caption{Vortex tangle structures emerging in the gas for the two different extremes of initial conditions, $\alpha=2.5$, at time $t=2424\,\tau$ (left) and $\alpha=10$, at time $t=606\,\tau$ (right). 
 }
 \label{fig:Phase}
 \end{figure}

\textit{Defect formation and dilution.}
In \Fig{Phase} we show, for intermediate times, the three dimensional distribution of points where the density falls below $0.2\%$ of the average density $\overline{n}$, for the systems quenched with $\alpha=2.5$ and $\alpha=10$.
We filtered out modes with wavenumber larger than $k\xi = 0.45$ but in the hydrodynamic case the high-momentum fluctuations barely distort the figure.
The vortex tangles corroborate the findings of \cite{Berloff2002a} for both cases of $\alpha$.
However, a remarkable difference exists in the distribution of the phase angle $\varphi(\mathbf{x})$ of the Bose field as can be inferred from \Fig{IncompSpectrum1}.
While after weak quenches, the circular flow has strong longitudinal (compressible) fluctuations, in the evolution passing the NTFP macroscopic quantized vortical flow and distinct Vinen tangles  \cite{Volovik2004a} appear in the superfluid.

\begin{figure}
  \includegraphics[width=0.37\textwidth]{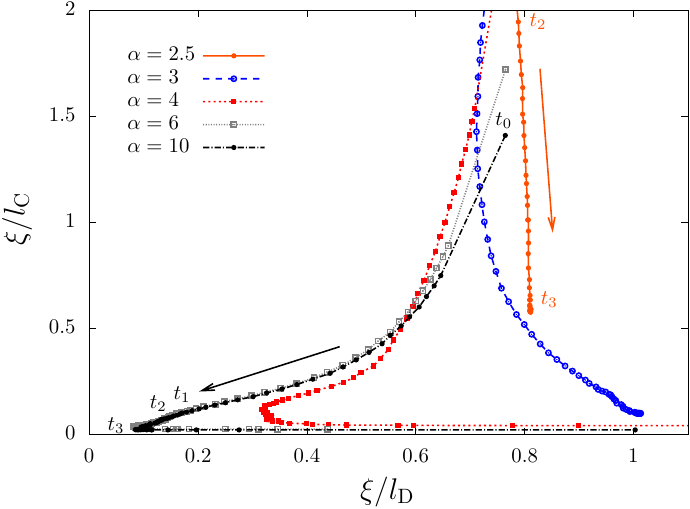}
  \caption{(Color online) 
  Trajectories of vortex states in the space of inverse coherence length $1/l_\mathrm{C}$ and inverse vortex-correlation length $1/l_\mathrm{D}$ for five different initial conditions. 
  Arrows are added to indicate the time direction. 
  The symbols are equally spaced on a logarithmic time scale except the step after $t_{0}$. 
  Times $t_0=0$, $t_1\approx 270\,\tau$, $t_2\approx 770\,\tau$ and $t_3\approx 6800\,\tau$ are marked for $\alpha=2.5, 10$, where $\tau=m\xi^{2}$.}
  \label{fig:PairingCoherence}
\end{figure}

%
The evolution of the vortex distribution, together with the phase coherence building up in the gas, eventually leading to a fully coherent BEC on a low-temperature background, allows to get a complementary understanding of the approach to and departure from the NTFP.
In \Fig{PairingCoherence}, we show the evolution of the system, again starting from the different initial quenches labelled by $\alpha$, in a reduced phase space defined by two different characteristic length scales, in units of the healing length $\xi$:
The coherence length $l_{\mathrm{C}}$ is a measure for the mean spatial fall-off of phase coherence in space.
We define it as the integral over the angle-averaged first-order coherence function $g^{(1)}(r)$ \cite{Supplement}.
The correlation length $l_{\mathrm{D}}$ measures the mean decay of  vorticity in the proximity of the vortex line defects in the system \cite{Supplement}.
For an isolated circular vortex ring, $l_{\mathrm{D}}$ is proportional to the diameter of the ring and thus has a similar relevance as the mean distance between vortices and anti-vortices in a 2D superfluid.
The trajectories of the quench-cooled Bose gas in the $(l_{\mathrm{D}},l_{\mathrm{C}})$-plane clearly exhibit the NTFP.
Lying in the lower left corner, it corresponds to a strongly coherent gas with a maximum mutual separation and minimum bending of the vortex filaments.
The NTFP corresponds ideally to a single vortex ring of maximum extent within the volume which is known to be a metastable configuration decaying only slowly via bending and sound generation \cite{Kozik2009a}. 
The marked time steps show the critical slowing down of the system approaching the NTFP, while colliding vortex rings are seen to reconnect and form larger rings.
After a long period in the vicinity of the fixed point the system eventually departs towards final thermal equilibrium.
This process corresponds to the shrinking of the last remaining vortex ring, by transferring energy to the incoherent sound excitations and particles to the condensate mode.
The NTFP sits at the crossroads of attractive and repulsive directions in our reduced phase space.

\textit{Summary.}
The process of Bose-Einstein condensation in a quench-cooled, dilute cold gas can show features of universal dynamics.
Provided a sufficiently strong cooling quench the condensing system passes by a partially attractive \emph{non-thermal fixed point} where it is critically slowed down.
The approach of the fixed point is marked by the appearance of incompressible flow around tangled vortex lines.
In this regime, particles can not be deposited quickly enough into the zero mode and form an excess population with a characteristic power-law fall-off within the low-energy modes. 
In contrast, slow, near-adiabatic condensation can exhibit the appearance of vortical motion which is, however, distorted by strong compressible sound excitations.
The critical slowing down of the phase coherence length and vortex distance provide smoking guns for the detection of the universal dynamics in experiment.
A complete characterization of NTFP in terms of a full set of critical exponents and thus universality classes, including anomalous dimensions, is most desirable, expanding the theory of critical phenomena far away from equilibrium. 
Understanding the possible different paths to a BEC is of fundamental interest way beyond the realm of ultracold gases, from the phenomenology of the solid state up to the highest energies, e.g., in heavy-ion collisions \cite{%
Berges:2008mr,Carrington:2010sz,Fukushima:2011nq,
Blaizot:2011xf,Berges:2012us}
or early-universe evolution \cite{Micha:2002ey,Berges:2008wm,
Berges:2008sr, 
Berges:2008mr,
Gasenzer:2011by}.

\textit{Acknowledgements}. 
We thank N. Berloff, J. Berges, J.~P.~Blaizot, S. Erne, M. Karl, L. McLerran, N. Philipp, D. Sexty, and B. Svistunov for useful discussions. 
Work supported by Deutsche Forschungsgemeinschaft (GA677/7,8), University of Heidelberg (FRONTIER), and Helmholtz Association (HA216/EMMI).


\newpage
\mbox{}
\newpage
\section{Supplementary Material}
\label{sec:SuppMat}
In the supplementary material we provide details of the methodology as well as additional results supporting our approach and conclusions.

%
\textit{Semiclassical simulations.}
Since the dynamics we are interested in exclusively affects the low-momentum, strongly populated field modes, we employ the so-called classical field method which yields, within numerical accuracy and the classical-wave approximation, exact results for the time-evolving observables~\cite{Blakie2008a2, Polkovnikov2010a2}. 
For this, initial field configurations $\phi(k,t_0)$ are sampled from Gaussian probability distributions and then propagated according to the classical equations of motion. 
At the end of the time evolution correlation functions are obtained from ensemble averages over the set of sampled paths. 
We study the dynamics of a dilute Bose gas by sampling statistically the classical equation of motion ($\hbar=1$)
\begin{equation}  \label{GPE}
   \i \del_t \phi(\mathbf{x},t)= \left[ -\frac{\nabla^2}{2m} + g|\phi(\mathbf{x},t)|^2 \right] \phi(\mathbf{x},t) .
\end{equation}
We consider a gas of $N$ atoms in a box of size $L^3$, with periodic boundary conditions and mean density $\overline{n}=N/L^3$. 
Lengths are measured in units of the healing length $\xi = (2mg\overline{n})^{-1/2}$ and time in units of $\tau=m\xi^{2}$.
Simulations were done on a cubic grid with $256^{3}$ points.

The initial field in momentum space, $ \phi(\mathbf{k},0) = \sqrt{n(\mathbf{k},0)} \mathrm{exp}\{ i \varphi (\mathbf{k},0) \} $,  is parametrized in terms of a randomly chosen phase $\varphi(\mathbf{k},0) \, \in \, [0,2\pi)$ and a density $n(\mathbf{k},0) = f(k)\nu_\mathbf{k}$, with $\nu_\mathbf{k}\ge0$ drawn from an exponential distribution $P(\nu_\mathbf{k})=\exp(-\nu_\mathbf{k})$ for each $\mathbf{k}$.
We choose the structure function
\begin{eqnarray}
\label{eq:structurefct}
 f(k) = \frac{f_{\alpha}}{k_{0}^{\alpha}+k^{\alpha}},
\end{eqnarray}
for different values of $\alpha$, with cutoff $(k_{0}\xi)^{\alpha}=0.2/0.44^{\alpha}$ and normalization $f_{\alpha}=400/0.44^{\alpha}$.
We compare results for a range of different cooling quenches defined by the power-laws $\alpha=2.5,\dots,10.0$, varying the total number between $N=10^{9}$ ($\alpha=2.5$) and $N=4.3\times10^{8}$ ($\alpha=10$).

\textit{Hydrodynamic decomposition.}
To interpret our results in the context of superfluid turbulence we analyse kinetic-energy spectra as in \cite{Nore1997a2}. 
The kinetic energy $E_{\mathrm{kin}}= \int \mathrm{d}\mathbf{x} \, \langle |\nabla \phi(\mathbf{x},t)|^2\rangle/2$ is split, $E_{\mathrm{kin}} = E_{\mathrm{v}} + E_\mathrm{q}$, into the `classical'  component $E_\mathrm{v}= \int \mathrm{d}\mathbf{x} \, \langle |\sqrt{n}\mathbf{v}|^2 \rangle/2 $ with velocity $\mathbf{v}= \nabla \varphi/m$, and a `quantum-pressure' component $E_\mathrm{q}=\int \mathrm{d}\mathbf{x} \, \langle |\nabla \sqrt{n}|^2 \rangle/2 $. 
Radial particle spectra
\begin{eqnarray}
 n_{\delta}(k)= \frac{1}{2} \int \mathrm{d}\Omega \, \langle |\mathbf{w}_{\delta}(\mathbf{k})|^2 \rangle,\quad \delta=\mathrm{v},\mathrm{q}.
\end{eqnarray}
 in terms of the generalised velocities $\vector{w}_{\mathrm{v}}=\sqrt{n}\vector{v}$ and $\vector{w}_{\mathrm{q}}=\nabla\sqrt{n}$ are furthermore decomposed by $\mathbf{w}_{\mathrm{v}}=\mathbf{w}_{\mathrm{i}}+\mathbf{w}_{\mathrm{c}}$ into `incompressible' ($\nabla \cdot \mathbf{w}_\mathrm{i}=0$) and `compressible' ($\nabla \times \mathbf{w}_\mathrm{c}=0$) parts to distinguish vortical superfluid and sound excitations of the gas, respectively.
The incompressible component reflects transverse motion, i.e., the flow which changes perpendicularly to its direction.
It arises mainly from superfluid vortical flow.
The compressible component accounts for the longitudinal flow corresponding to sound wave excitations.
%

%
\begin{figure}[!t]
\flushleft
\ \ \includegraphics[width=0.41\textwidth]{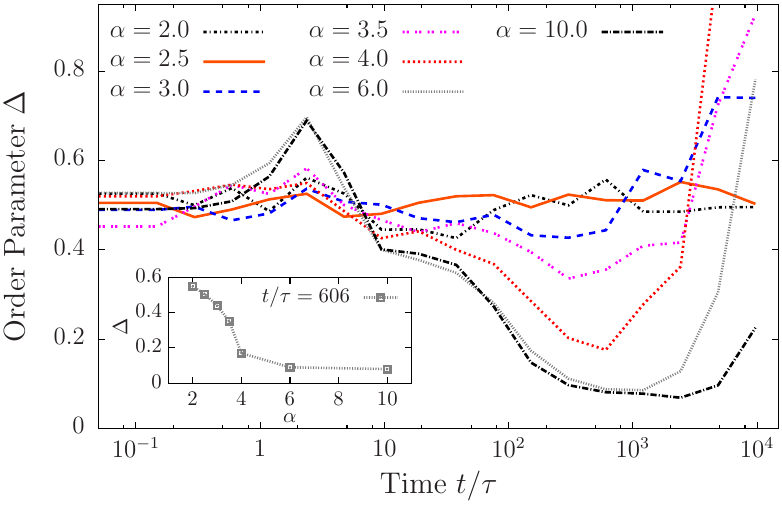}
\caption{(Color online) Evolution of the fraction $\Delta$ of integrated compressible to incompressible components below a momentum scale $k_{\lambda}\xi = 0.35$, for different initial conditions. 
Note the semi-log scale. 
For $\alpha\lesssim4$, $\Delta(t)$ stays approximately constant. 
For $\alpha \gtrsim 4$, $\Delta(t)$ approaches zero at intermediate times, signaling the non-thermal fixed point and a superfluid hydrodynamic condensation process.
Averages over $1...7$ runs. 
Inset: $\Delta(t=606\tau)$ as a function of $\alpha$. }
\label{fig:OrderParameter}
\end{figure}
%

 The evolution of the integrated fraction $\Delta(t) = N_c(k_{\lambda},t)/N_i(k_{\lambda},t)$, with $N_{\delta}(k_{\lambda},t)= \int_{|\mathbf{k}| < k_{\lambda}} \mathrm{d}\mathbf{k} \, n_{\delta}(\mathbf{k},t)$, of compressible to incompressible occupations below a momentum scale $k_{\lambda}\xi = 0.35$ is shown in \Fig{OrderParameter}. 
Initially, $\Delta \sim 0.5$. 
Starting from $\alpha \leq 3$, $\Delta$ stays approximately constant while for $\alpha > 3$, $\Delta$ decreases for $t \gtrsim 10^2 \tau$ towards zero before increasing again when thermal equilibrium is approached. 
The inset shows $\Delta(t=606)$ as a function of $\alpha$ in which one identifies a transition to a separation of the components, depending on the strength of the initial cooling quench.

\textit{Definition of the reduced phase space in terms of $l_\mathrm{C}$ and $l_\mathrm{D}$.}
The coherence length $l_\mathrm{C}$ is defined as the integral of the angle-averaged first-order coherence function 
\begin{eqnarray}
  g^{(1)}(r) 
  &=& \int \mathrm{d}\Omega \frac{\langle \phi^*(\mathbf{x})\phi(\mathbf{x}+\mathbf{r})\rangle}{\sqrt{\langle n(\mathbf{x})\rangle \langle n(\mathbf{x}+\mathbf{r})\rangle}},
  \\
  l_\mathrm{C} 
  &=& \int \mathrm{d}r \, g^{(1)}(r),
\end{eqnarray}
which becomes independent of $\mathbf{x}$ in the ensemble average.
In contrast to $r_\mathrm{coh} = \int \mathrm{d}r \, r^3 g^{(1)}(r) / \int \mathrm{d}r \, r^2 g^{(1)}(r)$ our choice of the coherence length $l_\mathrm{C}$ does not enlarge insignificant contributions at large r.

The vortex-correlation length is defined in terms of the vorticity. 
The curl of the velocity field $\mathbf{u}(\mathbf{x}) = \mathbf{rot} \, \mathbf{v}(\mathbf{x}) = \nabla\times\nabla \varphi(\mathbf{x}) / m$ vanishes except at the positions of topological defects. 
At these phase defects it yields the quantization and direction of the vortex line. 
Based on the vorticity $\mathbf{u}(\mathbf{x})$, we define angle-averaged correlations of the vorticity as
\begin{eqnarray}
  C(\mathbf{x},r) = \int \mathrm{d}\Omega_{\mathbf{r}} \, \mathbf{u}(\mathbf{x}) \mathbf{u}(\mathbf{x}+\mathbf{r}).
\end{eqnarray}
For small longitudinal distances $r$ along the vortex filament the vorticity $\mathbf{u}$ points in approximately the same direction as at $\mathbf{x}$, rendering $C(\mathbf{x},r)$ positive. 
In general, however, the vorticity shrinks with growing $r$, and the first transverse zero $r_0(\mathbf{x})$ marks the smallest distance, at which the vorticity points predominantly in the opposite direction, due to the curving of the same vortex line or the proximity of other vortex lines. 
$r_0(\mathbf{x})$ has therefore a similar meaning as the pair distance of anti-circulating vortices in two-dimensional gases and works equally well for vortex tangles and vortex rings. 
$l_\mathrm{D}$ is defined as the average of the transitions through zero,
\begin{eqnarray}
  l_\mathrm{D} = \mathcal{N}^{-1} \int_\mathcal{C} \mathrm{d} l \, r_0(l),
\end{eqnarray}
%
\begin{figure}[!t]
\begin{center}
  \includegraphics[width=0.4\textwidth]{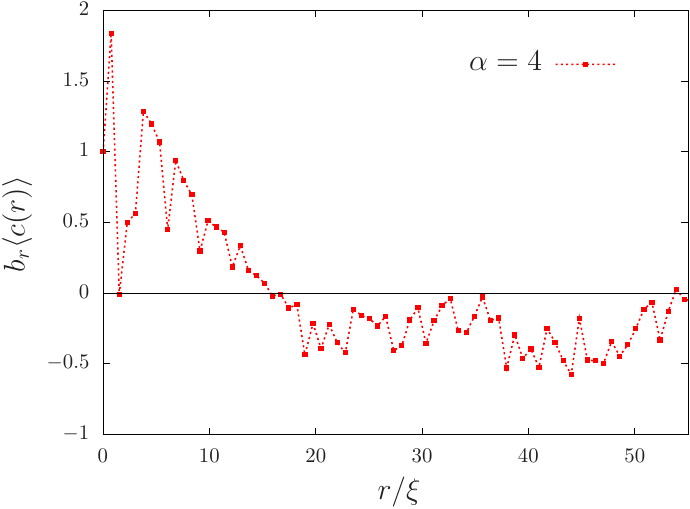}
\end{center}
  \caption{(Color online) 
  Ensemble averaged vortex correlations obtained from an initial state with $\alpha = 4$ at $t=2604\,\tau$. The correlations have been multiplied with a binning factor $b_{r}$, being equal to the number of grid points contributing to the angle average, approaching $b_{r}=4\pi r^{2}$ in the continuum limit.  
  This improves the visibility of the different transitions through zero. The transition through zero at $r \approx 17\xi$ is due to a large vortex ring. The transition through zero at $r \approx 2\xi$ and the minima at $r \approx 6\xi$ and $r \approx 9\xi$ are caused by several smaller rings.}
  \label{fig:VortexCorrelations}
\end{figure}
%
taking the normalization integral over all vortex lines and with $\mathcal{N} = \int_\mathcal{C} \mathrm{d}l$. 
Accounting for the transition through zero for each $\mathbf{x}$ has the advantage, that for small and large vortex rings, $r_0$ is calculated, in effect, separately and averaged with a weight corresponding to the length of each ring. 
If one took the transition through zero of the averaged correlation function $c(r)=[\int \mathrm{d}^3 x \, C(\mathbf{x},r)]/\int  \mathrm{d}^3 x \, |\mathbf{u}(\mathbf{x})|^{2}$, the positive contribution of a large ring at some distance $r$ could be of a similar order as the negative contributions of smaller rings (see \Fig{VortexCorrelations}). 
If both contributions almost cancel each other, small fluctuations around zero could lead to discontinuous jumps of $r_0$ with time.


\end{document}